\documentclass[epj,nopacs,final]{svjour}
\usepackage{graphics}
\newcommand{\PRD}[3]       {Phys.\ Rev.~{\bf D#1} (#2) #3}
\newcommand{\PRL}[3]       {Phys.\ Rev.\ Lett.~{\bf #1} (#2) #3}

\newcommand{\EPJ}[3]       {Eur.\ Phys.\ J.~{\bf C#1} (#2) #3}

\begin{document}
\hugehead
\sloppy
\title{\boldmath Angular distributions of leptons from $J/\psi$'s produced \\
in 920 GeV fixed-target proton-nucleus collisions}
\author{
I.~Abt\inst{24}\and M.~Adams\inst{11}\and M.~Agari\inst{14}\and H.~Albrecht\inst{13}\and
A.~Aleksandrov\inst{30}\and V.~Amaral\inst{9}\and A.~Amorim\inst{9}\and
S.~J.~Aplin\inst{13}\and V.~Aushev\inst{17}\and Y.~Bagaturia\inst{13\and37}
% ...
}
\institute{ NIKHEF, 1009 DB Amsterdam, The Netherlands~$^{a}$ \and Department ECM,
Faculty of Physics, University of Barcelona, E-08028 Barcelona, Spain~$^{b}$ \and
Institute for High Energy Physics, Beijing 100039, P.R. China \and Institute of
Engineering Physics, Tsinghua University, Beijing 100084, P.R. China \and Institut f\"ur
Physik, Humboldt-Universit\"at zu Berlin, D-12489 Berlin, Germany~$^{c,d}$ \and
Dipartimento di Fisica dell' Universit\`{a} di Bologna and INFN Sezione di Bologna,
I-40126 Bologna, Italy \and also from Fondazione Giuseppe Occhialini, I-61034 Fossombrone
(Pesaro Urbino), Italy \and Department of Physics, University of Cincinnati, Cincinnati,
Ohio 45221, USA~$^{e}$ \and LIP Coimbra, P-3004-516 Coimbra, Portugal~$^{f}$ \and Niels
Bohr Institutet, DK 2100 Copenhagen, Denmark~$^{g}$ \and Institut f\"ur Physik,
Universit\"at Dortmund, D-44221 Dortmund, Germany~$^{d}$ \and Joint Institute for Nuclear
Research Dubna, 141980 Dubna, Moscow region, Russia \and DESY, D-22603 Hamburg, Germany
\and Max-Planck-Institut f\"ur Kernphysik, D-69117 Heidelberg, Germany~$^{d}$ \and
Physikalisches Institut, Universit\"at Heidelberg, D-69120 Heidelberg, Germany~$^{d}$
\and Department of Physics, University of Houston, Houston, TX 77204, USA~$^{e}$ \and
Institute for Nuclear Research, Ukrainian Academy of Science, 03680 Kiev, Ukraine~$^{h}$
\and J.~Stefan Institute, 1001 Ljubljana, Slovenia~$^{i}$ \and University of Ljubljana,
1001 Ljubljana, Slovenia \and University of California, Los Angeles, CA 90024, USA~$^{j}$
\and Lehrstuhl f\"ur Informatik V, Universit\"at Mannheim, D-68131 Mannheim, Germany \and
University of Maribor, 2000 Maribor, Slovenia \and Institute of Theoretical and
Experimental Physics, 117218 Moscow, Russia~$^{k}$ \and Max-Planck-Institut f\"ur Physik,
Werner-Heisenberg-Institut, D-80805 M\"unchen, Germany~$^{d}$ \and Dept. of Physics,
University of Oslo, N-0316 Oslo, Norway~$^{l}$ \and Fachbereich Physik, Universit\"at
Rostock, D-18051 Rostock, Germany~$^{d}$ \and Fachbereich Physik, Universit\"at Siegen,
D-57068 Siegen, Germany~$^{d}$ \and Institute for Nuclear Research, INRNE-BAS, Sofia,
Bulgaria \and Universiteit Utrecht/NIKHEF, 3584 CB Utrecht, The Netherlands~$^{a}$ \and
DESY, D-15738 Zeuthen, Germany \and Physik-Institut, Universit\"at Z\"urich, CH-8057
Z\"urich, Switzerland~$^{m}$ \and visitor from Dipartimento di Energetica dell'
Universit\`{a} di Firenze and INFN Sezione di Bologna, Italy \and visitor from
P.N.~Lebedev Physical Institute, 117924 Moscow B-333, Russia \and visitor from Moscow
Physical Engineering Institute, 115409 Moscow, Russia \and visitor from Moscow State
University, 119992 Moscow, Russia \and visitor from Institute for High Energy Physics,
Protvino, Russia \and visitor from High Energy Physics Institute, 380086 Tbilisi,
Georgia\\ \vspace{1mm} \\
$^\dagger${\it deceased} \\
$^{a}$ supported by the Foundation for Fundamental Research on Matter (FOM), 3502 GA Utrecht, The Netherlands \\
$^{b}$ supported by the CICYT contract AEN99-0483 \\
$^{c}$ supported by the German Research Foundation, Graduate College GRK 271/3 \\
$^{d}$ supported by the Bundesministerium f\"ur Bildung und Forschung, FRG, under contract numbers 05-7BU35I, 05-7DO55P, 05-HB1HRA, 05-HB1KHA, 05-HB1PEA, 05-HB1PSA, 05-HB1VHA, 05-HB9HRA, 05-7HD15I, 05-7MP25I, 05-7SI75I \\
$^{e}$ supported by the U.S. Department of Energy (DOE) \\
$^{f}$ supported by the Portuguese Funda\c c\~ao para a Ci\^encia e Tecnologia under the program POCTI \\
$^{g}$ supported by the Danish Natural Science Research Council \\
$^{h}$ supported by the National Academy of Science and the Ministry of Education and Science of Ukraine \\
$^{i}$ supported by the Ministry of Education, Science and Sport of the Republic of Slovenia under contracts number P1-135 and J1-6584-0106 \\
$^{j}$ supported by the U.S. National Science Foundation Grant PHY-9986703 \\
$^{k}$ supported by the Russian Ministry of Education and Science, grant SS-1722.2003.2, and the BMBF via the Max Planck Research Award \\
$^{l}$ supported by the Norwegian Research Council \\
$^{m}$ supported by the Swiss National Science Foundation \\
\vspace{1mm} \\
\emph{Many thanks to Antonio Vitale (1943-2008)} }
\date{Received: date / Revised version: date}
%
\abstract{ A study of the angular distributions of leptons from decays of $J/\psi$'s
produced in p-C and p-W collisions at $\sqrt{s}=41.6$~GeV has been performed in the
$J/\psi$ Feynman-$x$ region $-0.34 < x_F < 0.14$ and for $J/\psi$ transverse momenta up
to $5.4$~GeV$/c$. The data were collected by the HERA-B experiment at the HERA proton
ring of the DESY laboratory. The results, based on a clean selection of $2.3 \cdot
10^{5}$ $J/\psi$'s reconstructed in both the $e^+ e^-$ and $\mu^+ \mu^-$ decay channels,
indicate that $J/\psi$'s are produced polarized. The magnitude of the effect is maximal
at low $p_T$. For $p_T > 1$~GeV/$c$ a significant dependence on the reference frame is
found: the polar anisotropy is more pronounced in the Collins-Soper frame and almost
vanishes in the helicity frame, where, instead, a significant azimuthal anisotropy
arises.
\PACS{
      {PACS-key}{discribing text of that key}   \and
      {PACS-key}{discribing text of that key}
     } 
} 
\authorrunning{HERA-B Coll.}
\titlerunning{Decay angular distributions of the $J/\psi$'s produced \\
in 920 GeV fixed-target proton-nucleus collisions}
\maketitle

\section{Introduction} \label{sec:introduction}
This paper presents a new measurement of the angular distribution of leptons from
$J/\psi$'s produced inclusively in proton-nucleus collisions at centre-of-mass energy
$\sqrt{s} = 41.6$~GeV. The data were collected by the DESY experiment HERA-B and covered
the kinematic ranges $-0.34 < x_F < 0.14$ in the Feynman-$x$ variable and $0 < p_T <
5.4$~GeV$/c$ in transverse momentum. In this domain, the average fraction of $J/\psi$
mesons coming from $\chi_c$ and $\psi^{\prime}$ decays has been determined as $\sim
27\%$~\cite{psip_HERAB,chic_HERAB}. Most previous analyses were based on the choice of
one specific definition of the polarization frame and were limited to the measurement of
the polar angle distribution, from which the so-called ``polarization'' parameter is
extracted. The present measurement includes for the first time a systematic comparison of
the results obtained for the \emph{full} decay angular distribution in three different
reference frames -- and significant differences are found between them. Some of the
results are presented separately for the two target materials used in the experiment
(carbon, $A=12$, and tungsten, $A=184$), leaving open the possibility that the nuclear
medium may affect the observed decay kinematics (for example as a consequence of a
varying mixture of $J/\psi$'s from decays of heavier charmonium states and direct
$J/\psi$'s). The analysis is based on almost the same sample used in the measurement of
the $J/\psi$ kinematic distributions described in \cite{jpsi_herab}, with a total of
about 83000 and 143000 $J/\psi$'s reconstructed, respectively, in the dimuon and
dielectron decay channels (excluding only the small fraction of data collected with
titanium). The reader is referred to that paper for a description of detector, data
taking, trigger, selection criteria and Monte Carlo simulation.

The next section explains the definitions and conventions used in the measurement
(Sect.~\ref{sec:definitions}). The results are presented in Sect.~\ref{sec:results} and
discussed in the conclusions (Sect.~\ref{sec:conclusions}).

\begin{figure*}
\centering \resizebox{0.79\textwidth}{!}{ \includegraphics{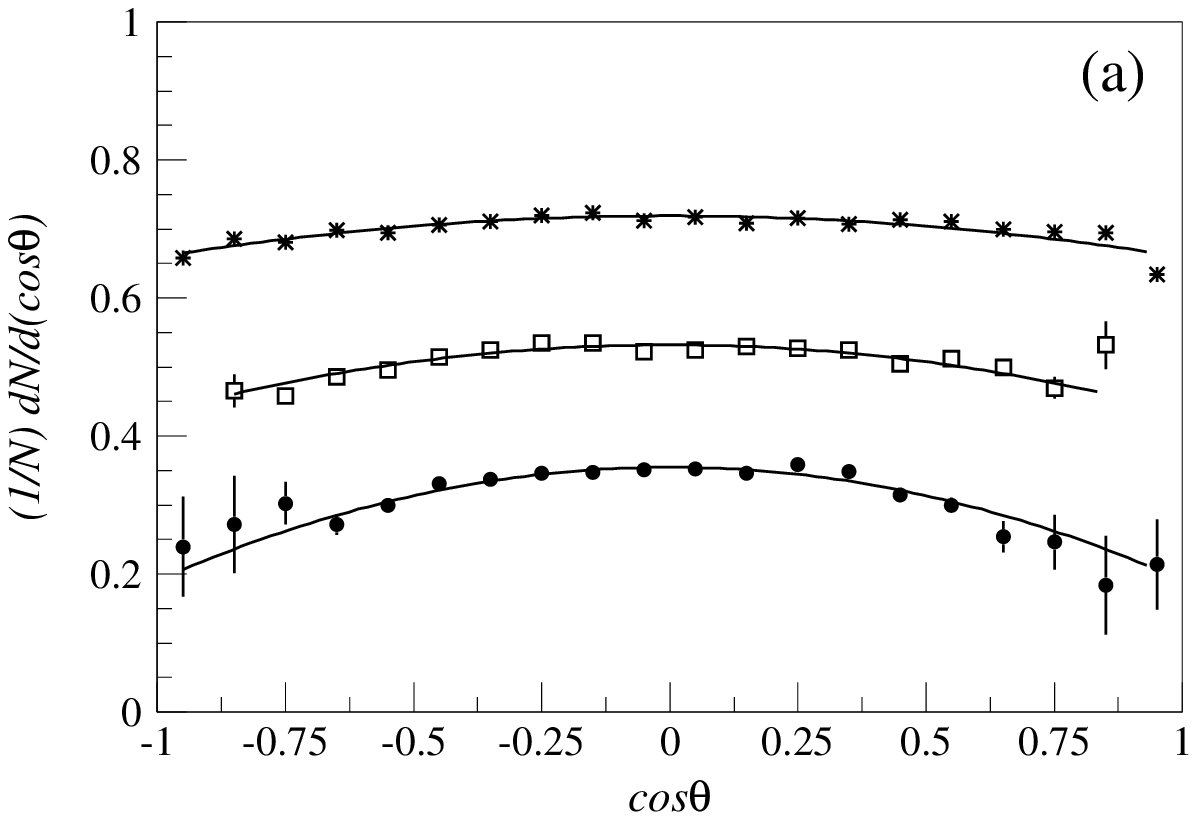}
                                          \includegraphics{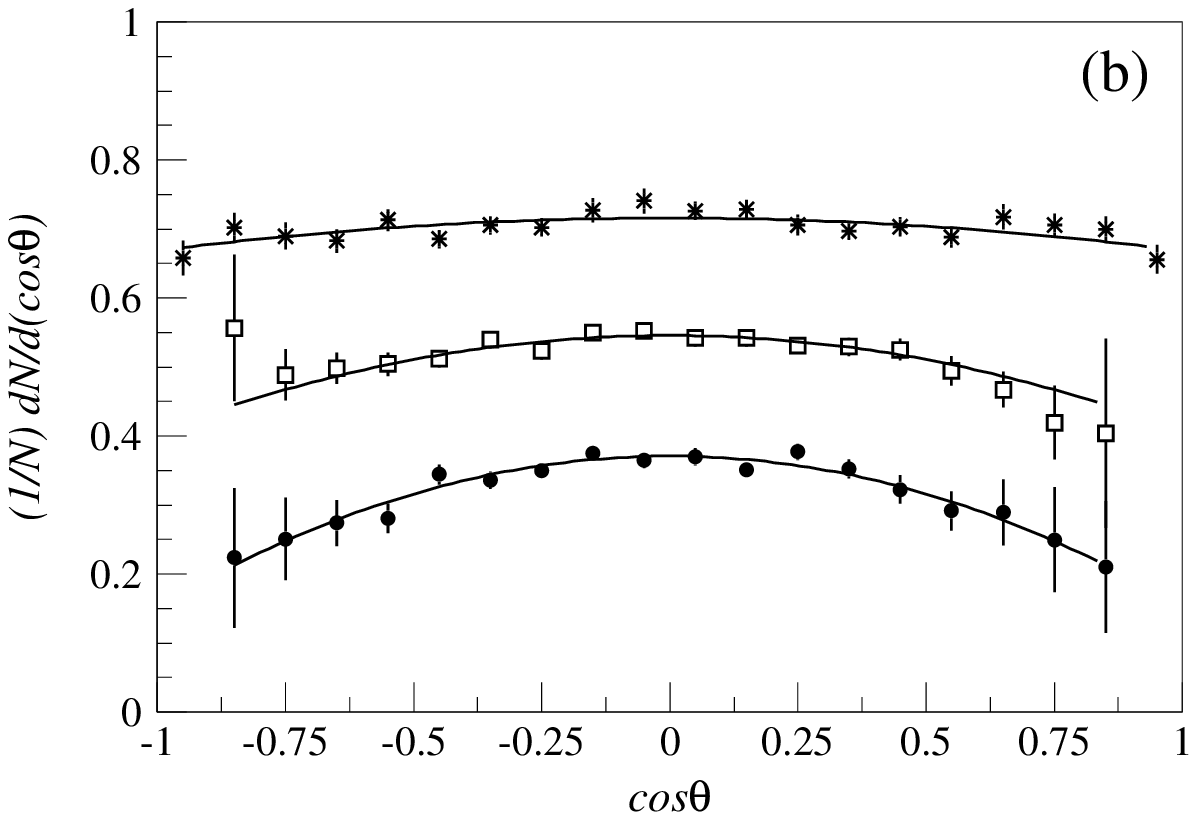} }
\centering \resizebox{0.79\textwidth}{!}{ \includegraphics{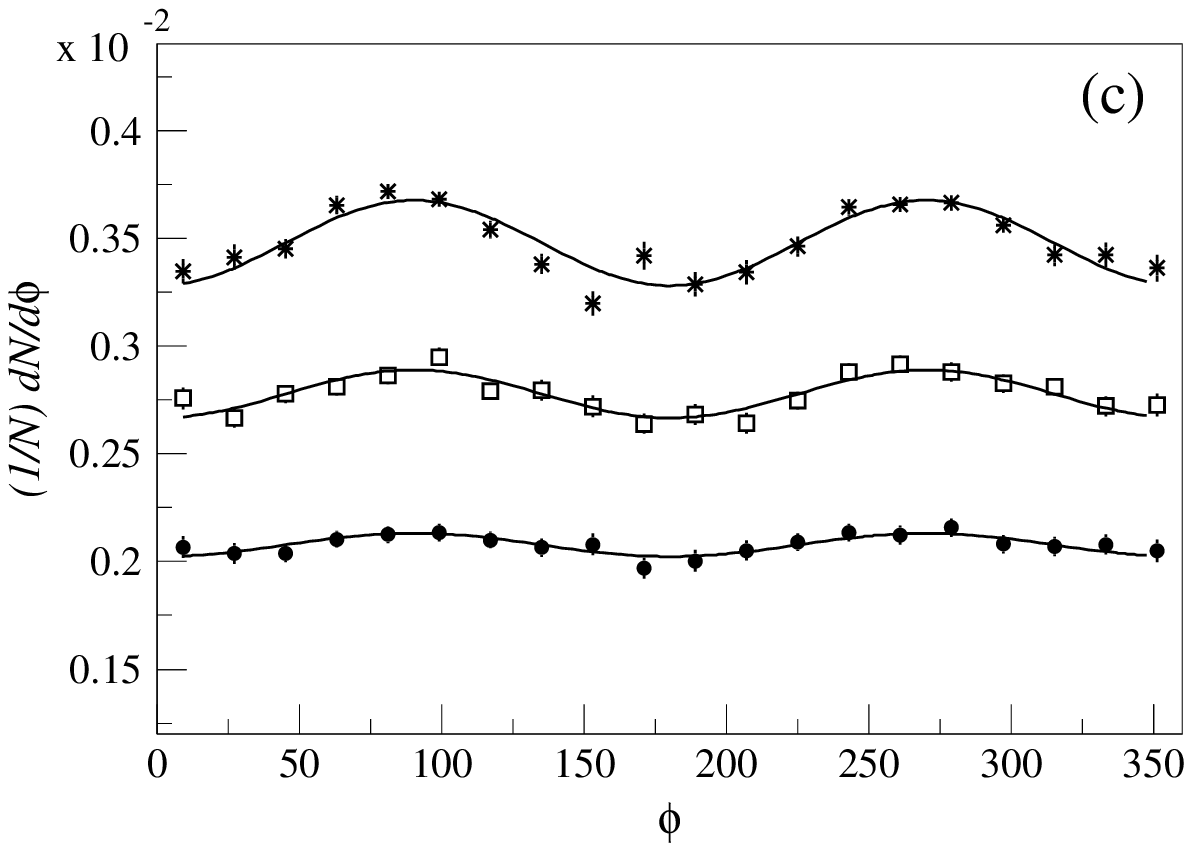}
                                          \includegraphics{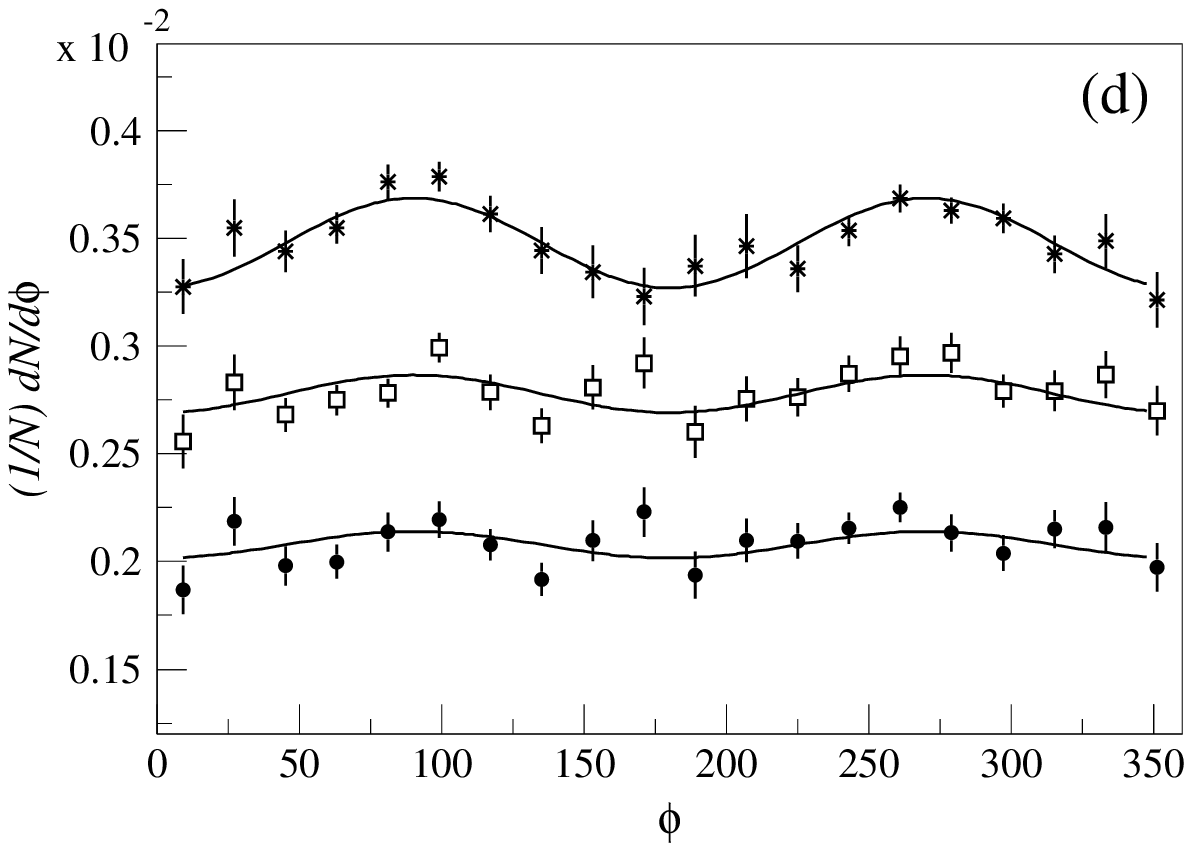} }
\centering \resizebox{0.79\textwidth}{!}{ \includegraphics{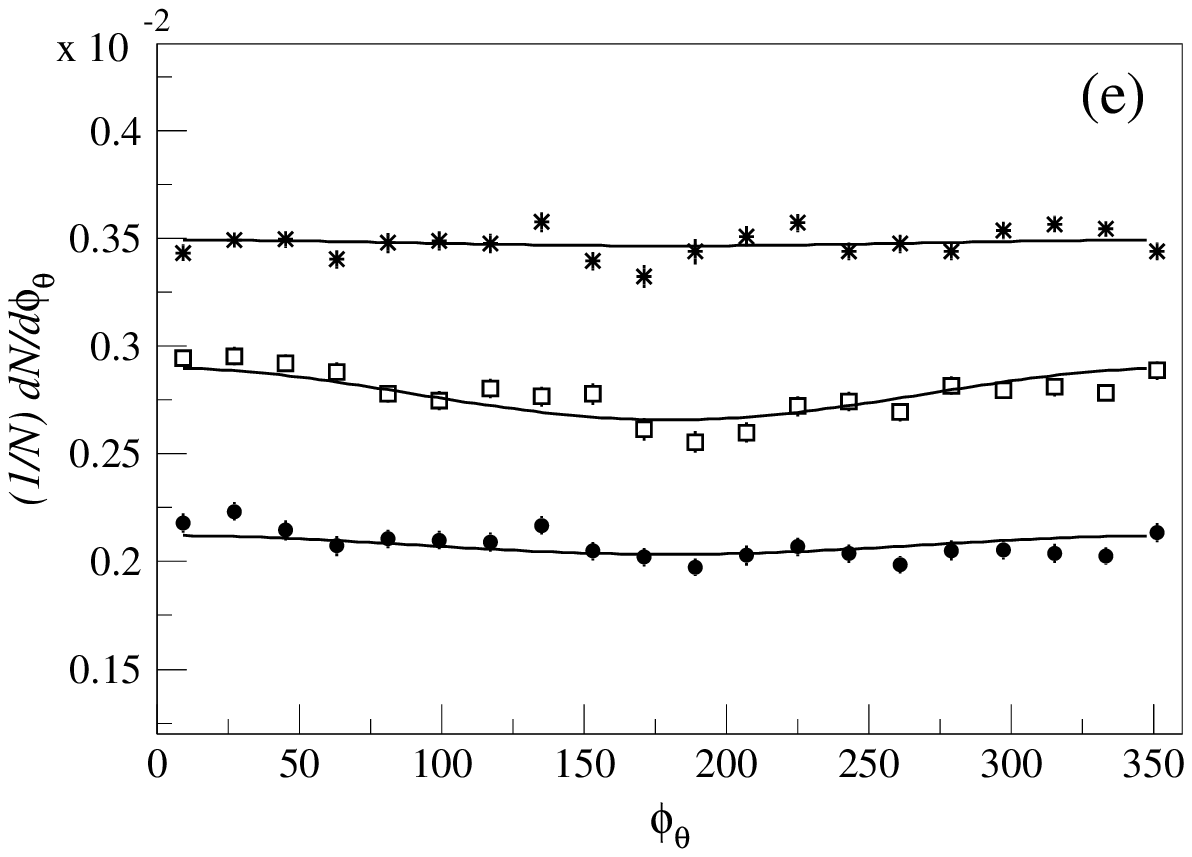}
                                          \includegraphics{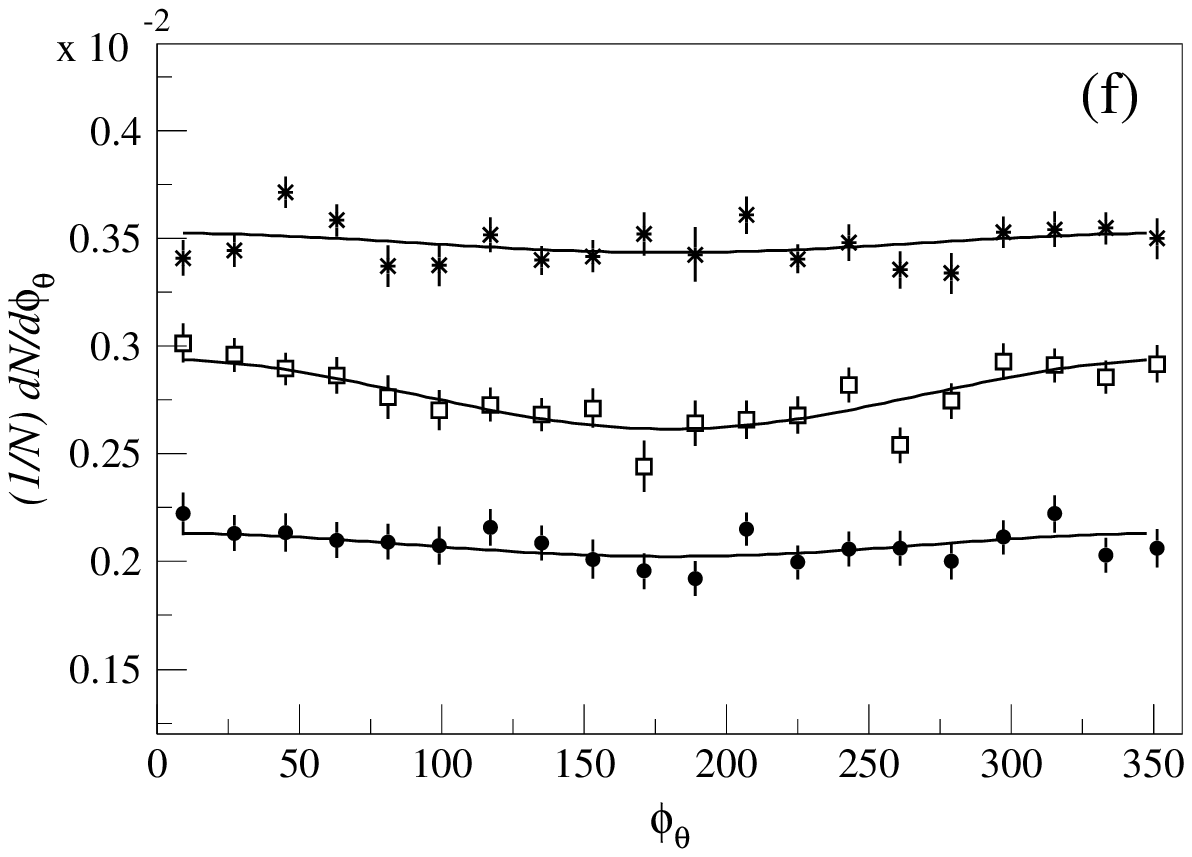} }
\caption{Efficiency-corrected distributions of the angular variables $\cos \theta$ (a,
b), $\phi$ (c, d) and $\phi_{\theta}$ (e, f) measured in the muon (left column) and
electron (right column) decay channels of the $J/\psi$ over the whole visible phase
space. The results obtained in the Collins-Soper, Gottfried-Jackson and helicity frames
are represented, respectively, by black circles, white squares and asterisks. The errors
are only statistical. For a better visualizations, the Collins-Soper and helicity
distributions are displayed with their actual values shifted by a constant. The
distributions are fitted with the curves of Eqs.~\ref{eq:costhdistr}, \ref{eq:phidistr}
and \ref{eq:phithdistr}.} \label{fig:acceptances}
\end{figure*}

\begin{table*}
\centering \caption{Output parameters obtained by fitting the distributions shown in
Fig.~\ref{fig:acceptances} with the curves of Eqs.~\ref{eq:costhdistr}, \ref{eq:phidistr}
and \ref{eq:phithdistr}. The errors in the parameters reflect only statistical
uncertainties of the distributions. The errors are correlated from one frame to another.
The systematic uncertainties for the average values are in all frames of the order of
$0.05$ for $\lambda_\theta$, $0.02$ for $\lambda_\phi$ and $0.015$ for $\lambda_{\theta
\phi}$.} \label{tab:accepance}
\begin{tabular}{cccc}
\hline\noalign{\smallskip}
Frame/channel & $\lambda_\theta$ & $\lambda_\phi$ & $\lambda_{\theta \phi}$  \\
\noalign{\smallskip}\hline\noalign{\smallskip}
CS/$\mu^+ \mu^-$  & $-0.296 \pm 0.029$ & $-0.0194 \pm  0.0051$  & $0.0158 \pm  0.0049$ \\
CS/$e^+ e^-$      & $-0.383 \pm 0.061$ & $-0.022  \pm  0.011$   & $0.0195 \pm  0.0096$  \\
CS/avg.           & $-0.313 \pm 0.026$ & $-0.0199 \pm 0.0046$   & $0.0168 \pm 0.0043$  \\
\noalign{\smallskip}\hline\noalign{\smallskip}
GJ/$\mu^+ \mu^-$  & $-0.185 \pm  0.021$ & $-0.0400 \pm  0.0051$ & $0.0433 \pm  0.0051$ \\
GJ/$e^+ e^-$      & $-0.256 \pm  0.051$ & $ -0.031 \pm  0.011$  & $0.058 \pm  0.010$  \\
GJ/avg.           & $-0.195 \pm  0.019$ & $-0.0385 \pm 0.0046$  & $0.0463 \pm 0.0045$  \\
\noalign{\smallskip}\hline\noalign{\smallskip}
HX/$\mu^+ \mu^-$  & $-0.115 \pm  0.012$ & $-0.0714 \pm  0.0055$ & $0.0049 \pm  0.0049$ \\
HX/$e^+ e^-$      & $-0.092 \pm  0.027$ & $ -0.075 \pm  0.012$  & $0.0161 \pm  0.0094$ \\
HX/avg.           & $-0.111 \pm  0.011$ & $-0.0720 \pm 0.0050$  & $0.0073 \pm 0.0043$ \\
\noalign{\smallskip}\hline
\end{tabular}
\end{table*}

\begin{figure*}
\centering \resizebox{1.0\textwidth}{!}{ \includegraphics{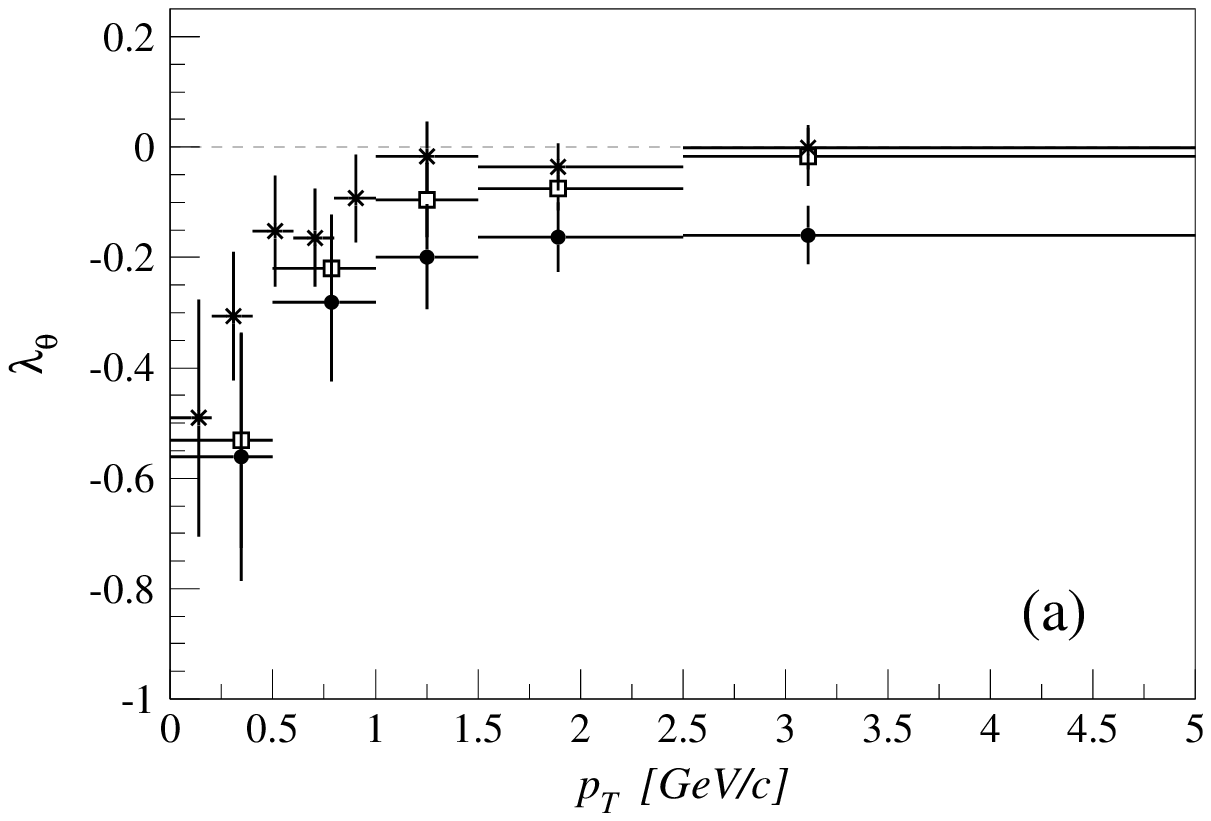}
                                          \includegraphics{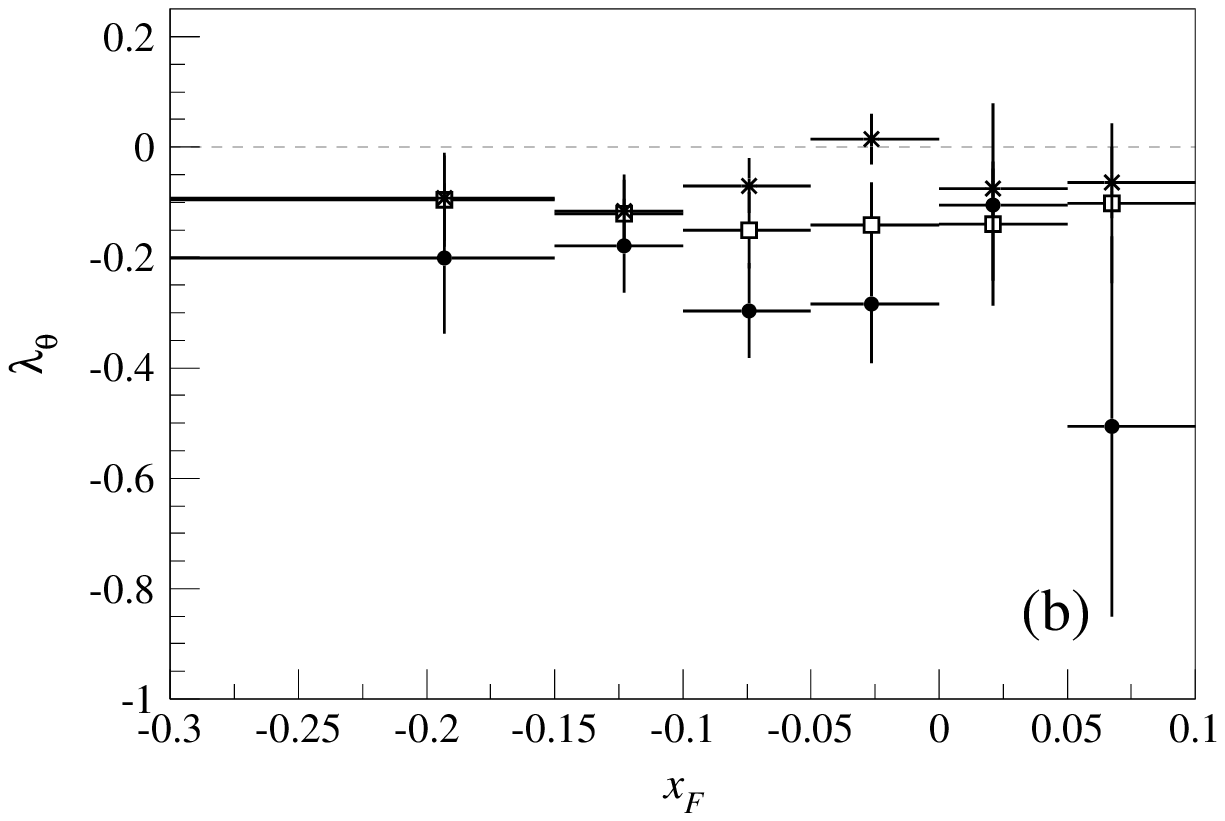} }
\centering \resizebox{1.0\textwidth}{!}{ \includegraphics{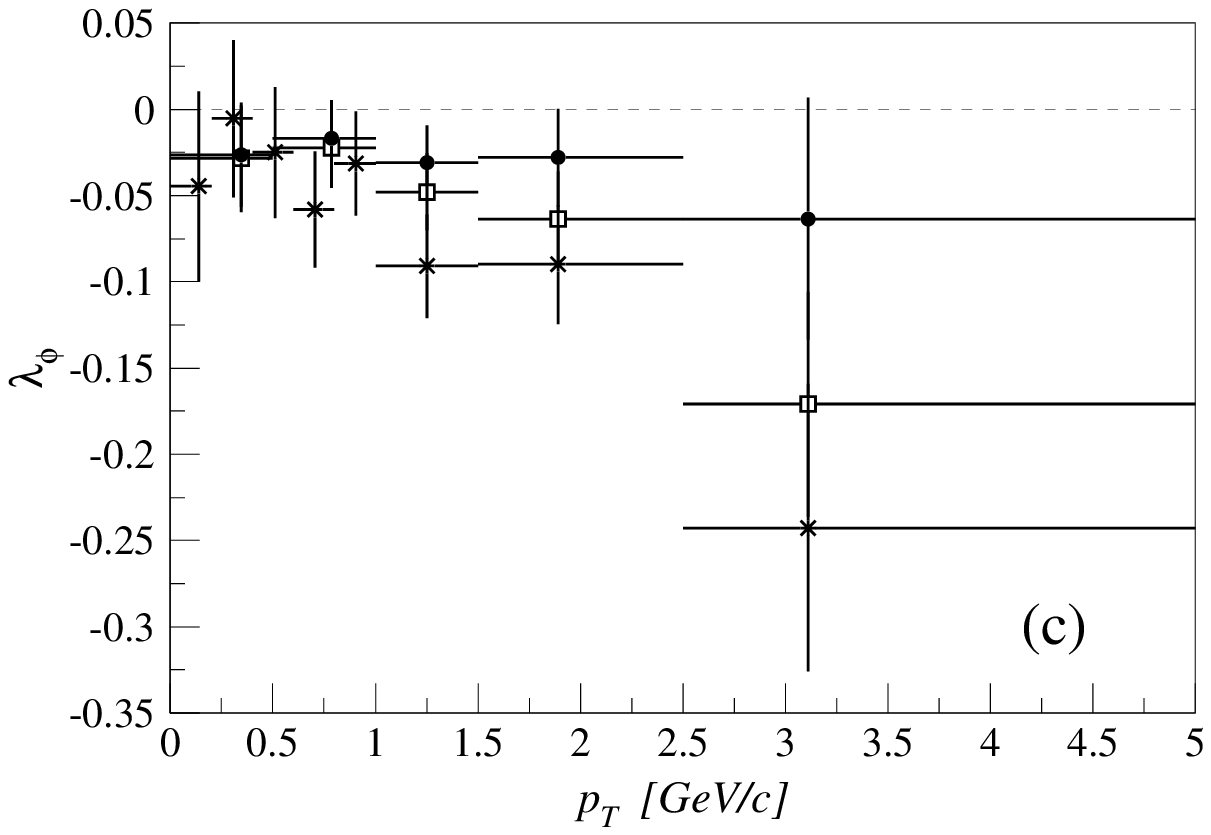}
                                          \includegraphics{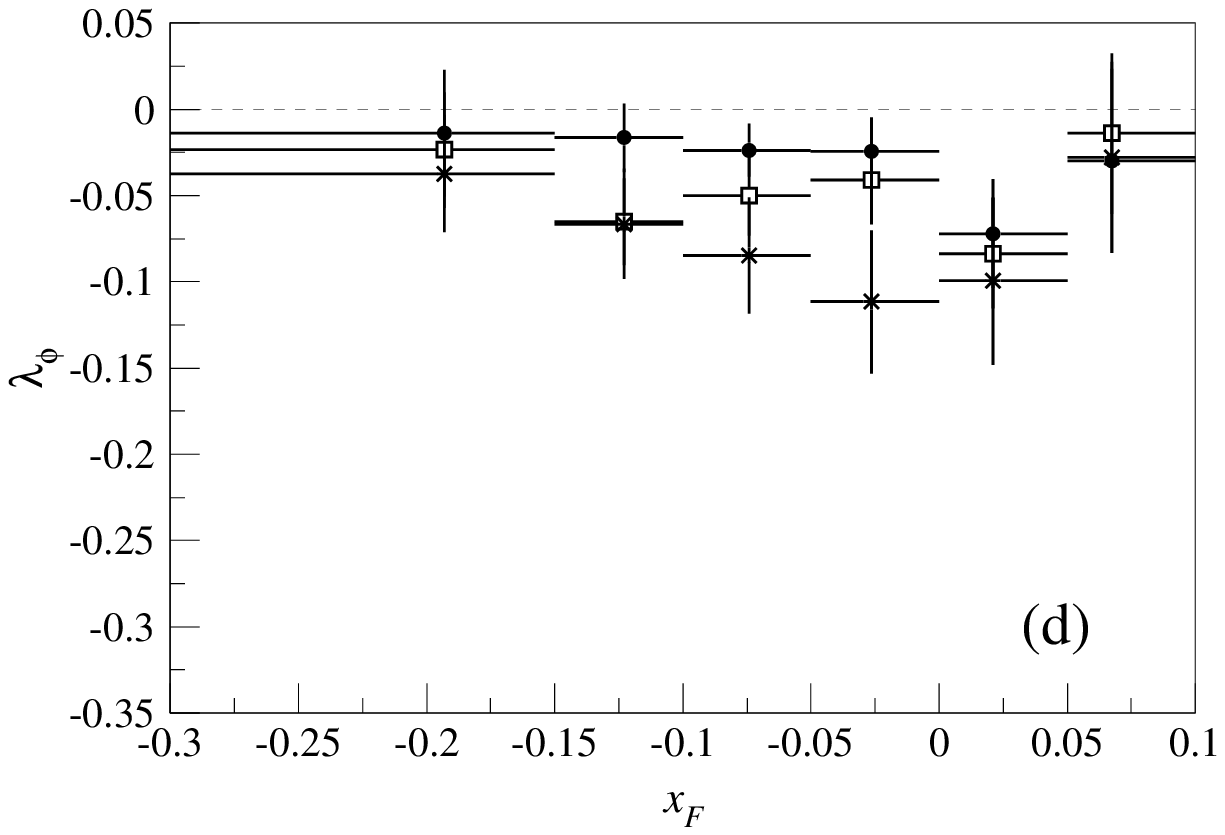} }
\centering \resizebox{1.0\textwidth}{!}{ \includegraphics{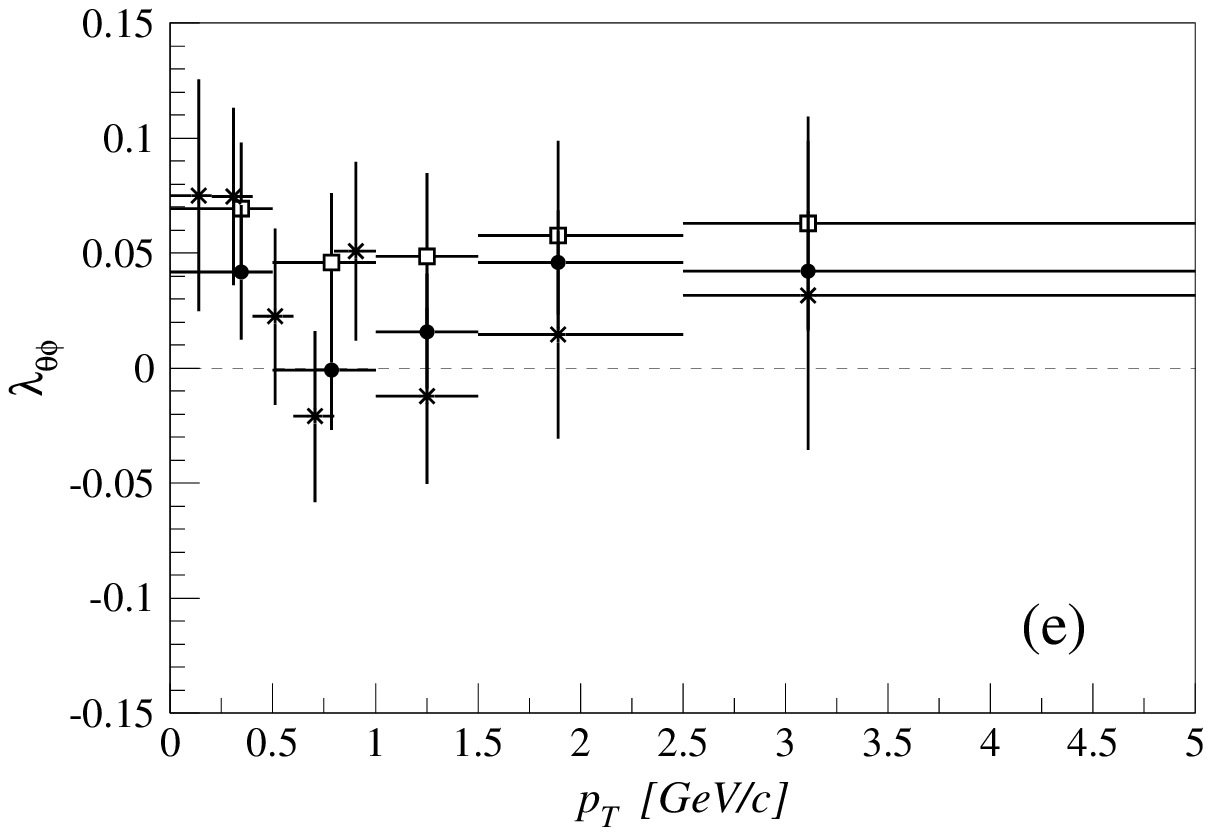}
                                          \includegraphics{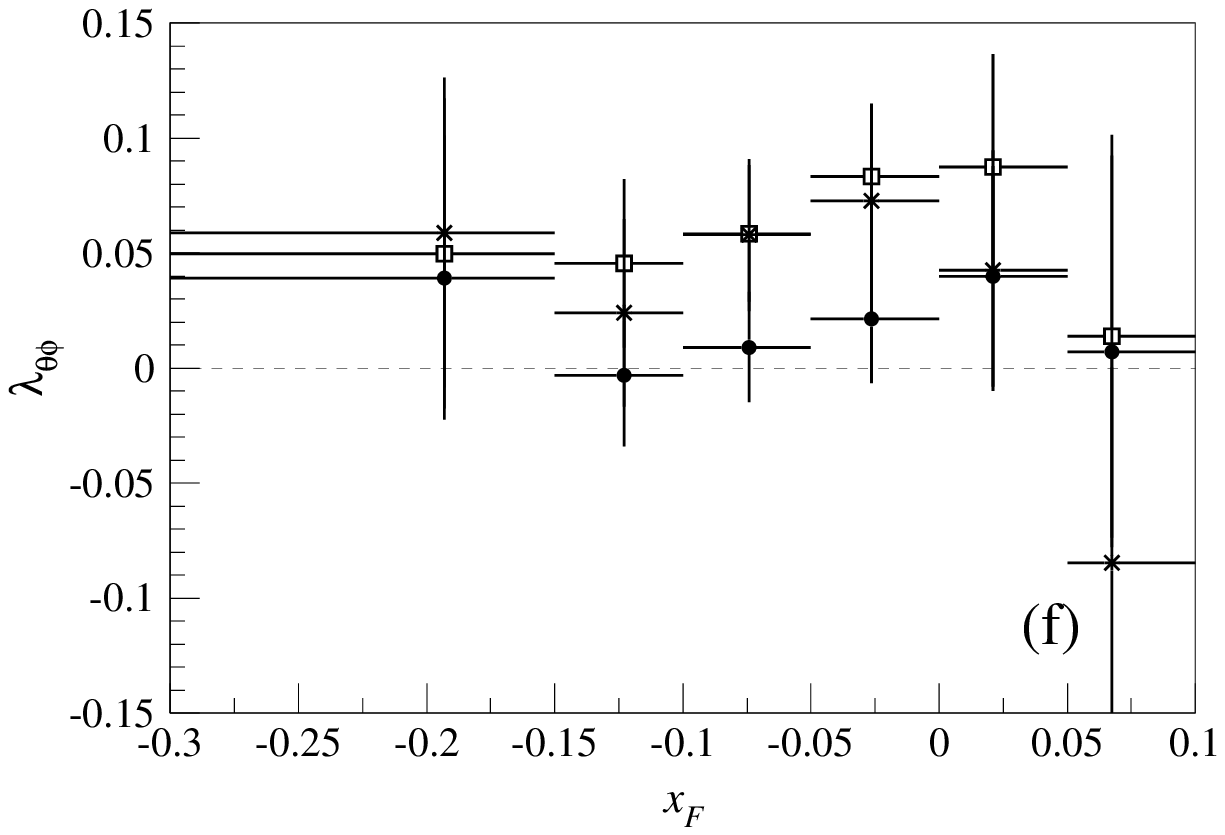} }
\caption{The parameters $\lambda_\theta$ (a, b), $\lambda_\phi$ (c, d) and
$\lambda_{\theta \phi}$ (e, f) measured as functions of the average reconstructed $p_T$
(left column) and $x_F$ (right column). The results obtained in the Collins-Soper,
Gottfried-Jackson and helicity frames are represented, respectively, by black circles,
white squares and asterisks. The vertical errors bars represent quadratic sums of
statistical and systematic uncertainties. The horizontal bars indicate the adopted
binning.} \label{fig:results}
\end{figure*}

\section{Definitions} \label{sec:definitions}

In any chosen reference system, the most general (parity conserving) form of the
two-lepton decay angular distribution of a $J/\psi$ is
\begin{eqnarray}
\frac{dN}{d(\cos\theta) \ d\phi} \; \propto \;  \; 1  & + & \lambda_\theta
\cos^2\theta \; + \; \lambda_\phi \sin^2\theta \cos (2 \phi) \nonumber \\
& + & \lambda_{\theta \phi} \sin (2 \theta) \cos \phi, \label{eq:angdistr}
\end{eqnarray}
where $\theta$ is the angle between the direction of the positive lepton and a chosen
polarization axis in the $J/\psi$ rest frame. $\phi$, the corresponding azimuthal angle,
is defined with respect to the plane of the colliding hadrons; $\lambda_{\theta}$,
$\lambda_{\theta \phi}$ and $\lambda_{\phi}$ are the quantities to be
measured.\footnote{The sign of the measured $\lambda_{\theta \phi}$ depends on the chosen
orientation of the $y$ axis (the one perpendicular to the plane of the colliding
hadrons). In the convention adopted for the present analysis, the $y$ axis is oriented as
the vector product of the beam momentum and the target momentum, $\vec{p}_{\mathrm{beam}}
\times \vec{p}_{\mathrm{target}}$.} The parameter $\lambda_{\theta}$ is usually called
``polarization''; more generally, any observed deviation of at least one of the three
parameters from zero would be the indication of polarized $J/\psi$ production. Non-zero
values of $\lambda_{\phi}$ and $\lambda_{\theta \phi}$ indicate an azimuthal anisotropy
of the distribution, which may arise as a consequence of specific choices of the
reference frame.

The following different definitions of the polarization axis are commonly used in the
literature:
\begin{itemize}
\item the direction of the beam momentum (as seen in the $J/\psi$ rest
frame) in the so-called Gottfried-Jackson (GJ) frame~\cite{gott_jack};
\item the direction of the $J/\psi$ in the center-of-mass system
of the colliding hadrons (helicity frame, HX);
\item the bisector between the directions of the beam momentum and of the opposite
of the target momentum (as seen in the $J/\psi$ rest frame) in the Collins-Soper (CS)
frame~\cite{coll_sop}.
\end{itemize}
The results of the present analysis are presented in all of these frames.

The values of $\lambda_\theta$, $\lambda_\phi$ and $\lambda_{\theta \phi}$ are extracted
from the data by considering single-variable projected angular distributions. When
averaged over $\phi$ and $\cos \theta$, respectively,
the angular distribution of the decay leptons is reduced to the forms
\begin{equation}
\frac{dN}{d(\cos\theta)} \; \propto \; 1 + \lambda_\theta \cos^2\theta
\label{eq:costhdistr}
\end{equation}
and
\begin{equation}
\frac{dN}{d \phi} \; \propto \;  1 + {\frac{2 \lambda_\phi}{3 + \lambda_\theta}} \cos (2
\phi). \label{eq:phidistr}
\end{equation}
One possible way of determining the value of $\lambda_{\theta \phi}$ is to define the
variable $\phi_\theta$ as
\begin{equation}
\phi_\theta = \left\{ \begin{array}{rcl}
\phi - \frac{3}{4} \pi & \mbox{for} & \cos \theta < 0 \\
\phi - \frac{\pi}{4} & \mbox{for} & \cos \theta > 0 \\
\end{array}
\right.
\end{equation}
and measure the distribution
\begin{equation}
\frac{dN}{d \phi_\theta} \; \propto \; 1 + {\frac{\sqrt{2}\lambda_{\theta\phi}}{3 +
\lambda_\theta}} \cos \phi_\theta. \label{eq:phithdistr}
\end{equation}

\section{Results} \label{sec:results}

The efficiency-corrected single-variable angular distributions averaged over the accepted
phase space ($-0.34 < x_F < 0.14$, $0 < p_T < 5.4$~GeV$/c$) are shown in
Fig.~\ref{fig:acceptances} with statistical uncertainties only. The results are given
separately in the two decay channels ($\mu^+ \mu^-$, $e^+ e^-$) for carbon and tungsten
target data combined. As can be seen, the measured distributions follow the correct
symmetric and/or periodic behaviour expressed by the formulas in
Eqs.~\ref{eq:costhdistr}, \ref{eq:phidistr} and \ref{eq:phithdistr}, indicating a good
level of reliability of the acceptance and efficiency correction procedures; the
chi-square probabilities obtained by fitting the distributions are on average 15\% and
50\% in the muon and electron channels, respectively. The values of the output parameters
of the fits are listed in Table~\ref{tab:accepance}. The difference between the values
measured in the two channels is always less than $1.3\sigma$. All results presented
hereafter are averages of muon and electron measurements. We remark that a preliminary
value of $-0.35 \pm 0.04$ was used for the effective $\lambda_{\theta}$ (CS frame) in the
evaluation of systematic uncertainties of a previous HERA-B analysis~\cite{chic_HERAB};
such value deviates slightly, but well within errors, from the combined value in
Table~\ref{tab:accepance}.

The final results including the estimated systematic uncertainties are displayed in
Fig.~\ref{fig:results} as a function of the transverse momentum and Feynman-$x$ of the
$J/\psi$. As before, the two target data samples have been combined. The corresponding
numerical values are listed in Tables~\ref{tab:results_CS}, \ref{tab:results_GJ} and
\ref{tab:results_HX}, where $\langle p_T \rangle$ and $\langle x_F \rangle$ indicate
averages over the $J/\psi$'s reconstructed in a given bin. The bin boundaries are defined
by the following lists: $0, 0.5, 1.0, 1.5, 2.5, 5.4$~GeV$/c$ ($0, 0.2, 0.4, 0.6, 0.8,
1.0, 1.5, 2.5, 5.4$~GeV$/c$ for the helicity frame) for $p_T$ and $-0.34, -0.15, -0.10,
-0.05, 0, 0.05, 0.14$ for $x_F$. The systematic errors have been evaluated with the
procedure already described in our report on the measurement of the $J/\psi$ kinematic
distributions~\cite{jpsi_herab}, taking into account the impact of signal selection and
optimization, signal counting method, differences in acquisition conditions, and the
kinematics of the MC generation. Additional systematic tests consisted in fitting the
angular distributions excluding the angular ranges with the lowest efficiency.
Statistical and systematic errors are obviously correlated from frame to frame. The
systematic errors in different $p_T$/$x_F$ bins are partly correlated.

The results indicate an anisotropy of the $J/\psi$ decay angular distribution, visible in
either its polar or azimuthal projections (in the CS and HX frames, respectively), or in
both (GJ frame). Moreover, there is a definite hierarchy in the magnitudes of the
parameters $\lambda_\theta$ and $\lambda_\phi$. In particular, the polar anisotropy
($\lambda_\theta < 0$) increases when going from the HX to the CS frame, while the
azimuthal parameter $\lambda_\phi$ changes following a reversed order. Both parameters
have in-between magnitudes in the GJ frame. A kinematic dependence characterizes the
results. For example, the magnitude of the polarization parameter $\lambda_\theta$
increases with decreasing $p_T$. This low-$p_T$ effect is the same in the three frames --
as expected from the fact that all frames coincide at $p_T = 0$ -- whereas different
polarization magnitudes are measured at higher $p_T$, following the already mentioned
hierarchy.

\begin{figure*}
\centering \resizebox{1.0\textwidth}{!}{ \includegraphics{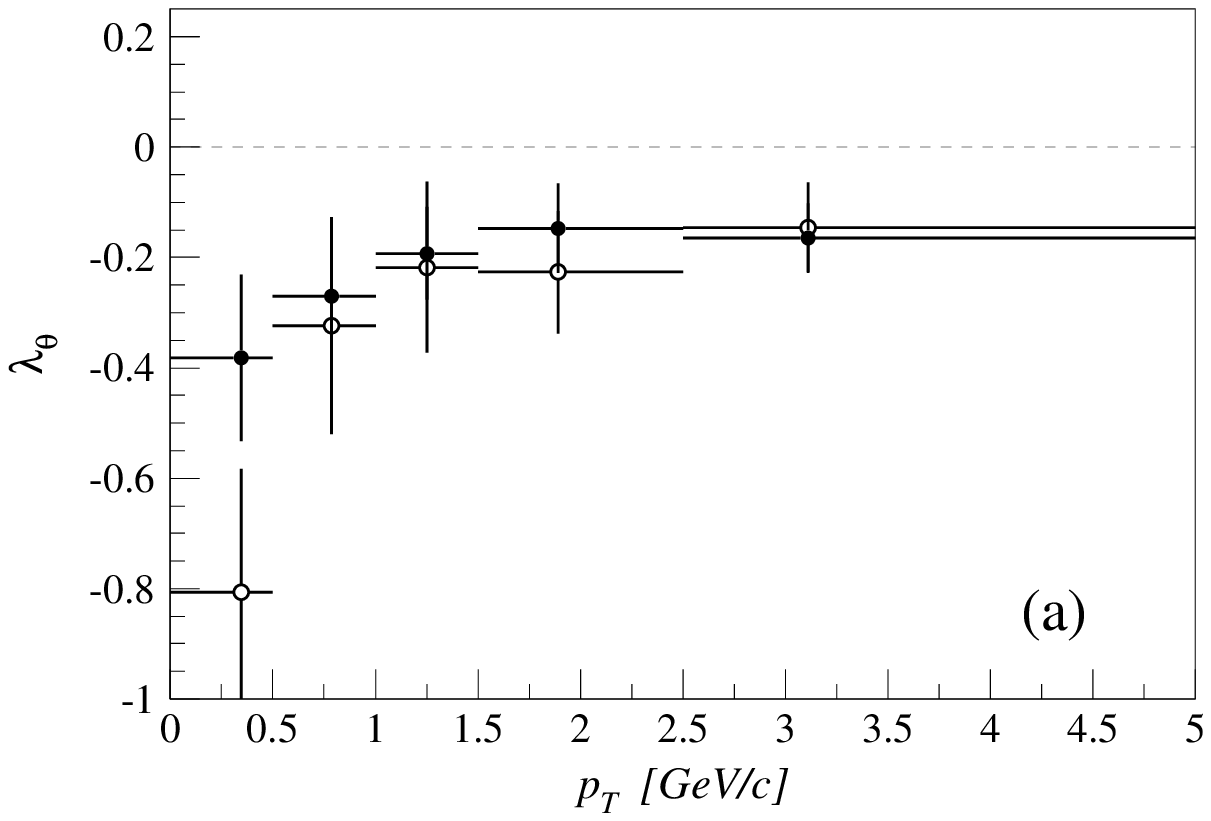}
                                          \includegraphics{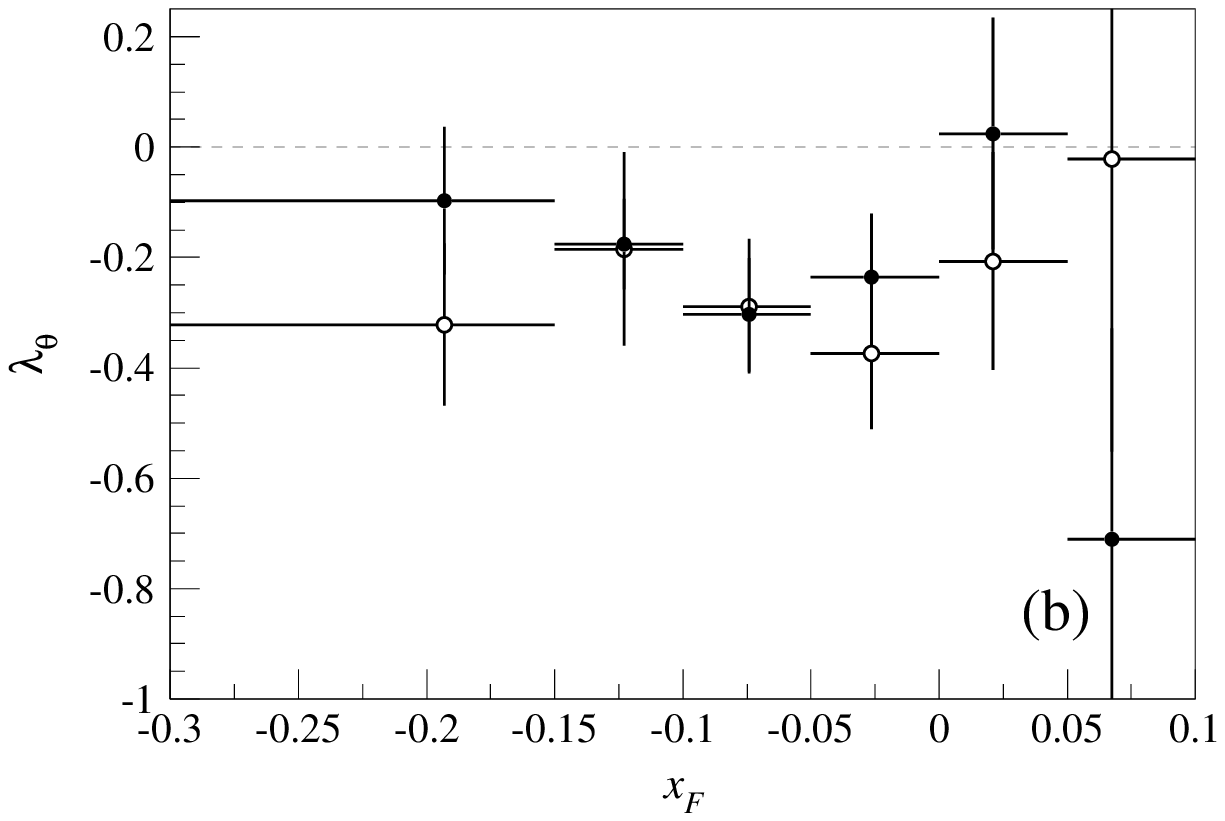} }
\caption{The parameter $\lambda_\theta$ measured in the Collins-Soper frame in carbon
(black points) and tungsten (white points) as functions of the average reconstructed
$p_T$ (a) and $x_F$ (b). The vertical error bars represent quadratic sums of statistical
and systematic uncertainties. The horizontal bars indicate the adopted binning.}
\label{fig:results_CW}
\end{figure*}

By performing a target-dependent analysis, generally small differences between the
polarization parameters measured with carbon and tungsten data have been found and taken
into account in the systematic errors of the already shown combined measurements.
However, the results for the parameter $\lambda_\theta$ show a slightly significant
target dependence in the lowest $p_T$ bin. The polarization measured in the CS frame as a
function of $p_T$ and $x_F$ is shown in Fig.~\ref{fig:results_CW} for the two target
materials; the numbers are listed in Table~\ref{tab:results_CW}. In the first $p_T$ bin,
carbon and tungsten results differ by about $3\sigma$ at the statistical level; the
significance of the difference is reduced to the $2\sigma$ level when also the systematic
errors are included in the comparison.

Since the largest polarization effects are seen at low $p_T$, special attention has been
devoted to the investigation of the systematic variations in the value of
$\lambda_\theta$ for $0<p_T<0.5$~GeV$/c$. In fact, the determination of the low-$p_T$
efficiency is especially sensitive to the Monte Carlo description of the detector region
close to the beam pipe. Various tests have been performed, which consisted in reducing
the geometrical acceptance around the beam pipe, changing the beam direction in the Monte
Carlo and testing different selection criteria for the low momentum leptons. As an
extreme check, the acceptance determination has been varied to reproduce the detector
conditions of acquisition periods not correlated to the considered sample. Final
systematic uncertainties of $\simeq 0.1$ and $\simeq 0.2$ have been evaluated for
$\lambda_\theta$ for the carbon and tungsten targets, respectively
(Table~\ref{tab:results_CW}).

\begin{table*}
\centering \caption{Values of the parameters $\lambda_\theta$, $\lambda_\phi$ and
$\lambda_{\theta \phi}$ measured in the Collins-Soper frame as functions of the average
reconstructed $p_T$ and $x_F$ for combined carbon and tungsten data. The errors are
statistical and systematic.} \label{tab:results_CS}
\begin{tabular}{cccc}
\hline\noalign{\smallskip}
$\langle p_T \rangle$ (GeV/$c$) & $\lambda_\theta$ & $\lambda_\phi$ & $\lambda_{\theta \phi}$  \\
\noalign{\smallskip}\hline\noalign{\smallskip}
 $0.35$  &  $-0.56 \pm 0.07 \pm 0.21$     &  $-0.026 \pm 0.013 \pm 0.027$  &  $0.042 \pm 0.020 \pm 0.022$  \\
 $0.79$  &  $-0.28 \pm 0.05 \pm 0.13$     &  $-0.017 \pm 0.010 \pm 0.020$  &  $-0.001 \pm 0.014 \pm 0.022$  \\
 $1.25$  &  $-0.199 \pm 0.046 \pm 0.083$  &  $-0.031 \pm 0.012 \pm 0.018$  &  $0.016 \pm 0.014 \pm 0.021$  \\
 $1.89$  &  $-0.164 \pm 0.049 \pm 0.039$  &  $-0.028 \pm 0.017 \pm 0.023$  &  $0.046 \pm 0.015 \pm 0.017$  \\
 $3.11$  &  $-0.159 \pm 0.040 \pm 0.034$  &  $-0.063 \pm 0.035 \pm 0.061$  &  $0.042 \pm 0.024 \pm 0.011$  \\
\noalign{\smallskip}\hline\noalign{\smallskip}
\noalign{\smallskip}\hline\noalign{\smallskip}
$\langle x_F \rangle$ & $\lambda_\theta$ & $\lambda_\phi$ & $\lambda_{\theta \phi}$  \\
\noalign{\smallskip}\hline\noalign{\smallskip}
$-0.193$  &  $-0.202 \pm 0.072 \pm 0.116$  &  $-0.014 \pm 0.019 \pm 0.032$  &  $0.039 \pm 0.023 \pm 0.057$  \\
$-0.123$  &  $-0.179 \pm 0.050 \pm 0.068$  &  $-0.017 \pm 0.014 \pm 0.014$  &  $-0.003 \pm 0.019 \pm 0.024$  \\
$-0.074$  &  $-0.296 \pm 0.052 \pm 0.068$  &  $-0.024 \pm 0.010 \pm 0.012$  &  $0.009 \pm 0.016 \pm 0.018$  \\
$-0.026$  &  $-0.284 \pm 0.051 \pm 0.094$  &  $-0.025 \pm 0.011 \pm 0.017$  &  $0.021 \pm 0.013 \pm 0.025$  \\
 $0.021$  &  $-0.10 \pm 0.11 \pm 0.15$     &  $-0.072 \pm 0.013 \pm 0.029$  &  $0.040 \pm 0.018 \pm 0.044$  \\
 $0.067$  &  $-0.51 \pm 0.27 \pm 0.22$     &  $-0.030 \pm 0.025 \pm 0.047$  &  $0.007 \pm 0.039 \pm 0.076$  \\
\noalign{\smallskip}\hline
\end{tabular}
\end{table*}

\begin{table*}
\centering \caption{Values of the parameters $\lambda_\theta$, $\lambda_\phi$ and
$\lambda_{\theta \phi}$ measured in the Gottfried-Jackson frame as functions of the
average reconstructed $p_T$ and $x_F$ for combined carbon and tungsten data. The errors
are statistical and systematic.} \label{tab:results_GJ}
\begin{tabular}{cccc}
\hline\noalign{\smallskip}
$\langle p_T \rangle$ (GeV/$c$) & $\lambda_\theta$ & $\lambda_\phi$ & $\lambda_{\theta \phi}$  \\
\noalign{\smallskip}\hline\noalign{\smallskip}
 $0.35$  &  $-0.53 \pm 0.06 \pm 0.19$     &  $-0.029 \pm 0.014 \pm 0.028$  &  $0.069 \pm 0.020 \pm 0.020$  \\
 $0.79$  &  $-0.219 \pm 0.033 \pm 0.091$  &  $-0.022 \pm 0.010 \pm 0.021$  &  $0.046 \pm 0.014 \pm 0.027$  \\
 $1.25$  &  $-0.096 \pm 0.033 \pm 0.059$  &  $-0.048 \pm 0.014 \pm 0.018$  &  $0.049 \pm 0.017 \pm 0.032$  \\
 $1.89$  &  $-0.075 \pm 0.025 \pm 0.031$  &  $-0.064 \pm 0.017 \pm 0.022$  &  $0.058 \pm 0.018 \pm 0.037$  \\
 $3.11$  &  $-0.018 \pm 0.041 \pm 0.033$  &  $-0.171 \pm 0.036 \pm 0.054$  &  $0.063 \pm 0.026 \pm 0.038$  \\
\noalign{\smallskip}\hline\noalign{\smallskip}
\noalign{\smallskip}\hline\noalign{\smallskip}
$\langle x_F \rangle$ & $\lambda_\theta$ & $\lambda_\phi$ & $\lambda_{\theta \phi}$  \\
\noalign{\smallskip}\hline\noalign{\smallskip}
$-0.193$  &  $-0.096 \pm 0.039 \pm 0.075$  &  $-0.023 \pm 0.021 \pm 0.026$  &  $0.050 \pm 0.024 \pm 0.063$  \\
$-0.123$  &  $-0.120 \pm 0.033 \pm 0.063$  &  $-0.065 \pm 0.015 \pm 0.020$  &  $0.046 \pm 0.020 \pm 0.031$  \\
$-0.074$  &  $-0.151 \pm 0.028 \pm 0.063$  &  $-0.050 \pm 0.012 \pm 0.020$  &  $0.059 \pm 0.018 \pm 0.024$  \\
$-0.026$  &  $-0.141 \pm 0.031 \pm 0.070$  &  $-0.041 \pm 0.011 \pm 0.023$  &  $0.083 \pm 0.014 \pm 0.029$  \\
 $0.021$  &  $-0.140 \pm 0.057 \pm 0.084$  &  $-0.084 \pm 0.013 \pm 0.029$  &  $0.088 \pm 0.019 \pm 0.045$  \\
 $0.067$  &  $-0.10 \pm 0.10 \pm 0.10$     &  $-0.014 \pm 0.026 \pm 0.039$  &  $0.014 \pm 0.049 \pm 0.072$  \\
\noalign{\smallskip}\hline
\end{tabular}
\end{table*}

\begin{table*}
\centering \caption{Values of the parameters $\lambda_\theta$, $\lambda_\phi$ and
$\lambda_{\theta \phi}$ measured in the helicity frame as functions of the average
reconstructed $p_T$ and $x_F$ for combined carbon and tungsten data. The errors are
statistical and systematic.} \label{tab:results_HX}
\begin{tabular}{cccc}
\hline\noalign{\smallskip}
$\langle p_T \rangle$ (GeV/$c$) & $\lambda_\theta$ & $\lambda_\phi$ & $\lambda_{\theta \phi}$  \\
\noalign{\smallskip}\hline\noalign{\smallskip}
 $0.14$  &  $-0.49 \pm 0.07 \pm 0.20$     &  $-0.045 \pm 0.029 \pm 0.047$  &  $0.075 \pm 0.043 \pm 0.026$  \\
 $0.31$  &  $-0.31 \pm 0.05 \pm 0.11$     &  $-0.005 \pm 0.021 \pm 0.041$  &  $0.075 \pm 0.028 \pm 0.027$  \\
 $0.51$  &  $-0.153 \pm 0.031 \pm 0.096$  &  $-0.025 \pm 0.017 \pm 0.034$  &  $0.022 \pm 0.026 \pm 0.028$  \\
 $0.71$  &  $-0.164 \pm 0.026 \pm 0.085$  &  $-0.058 \pm 0.018 \pm 0.029$  &  $-0.021 \pm 0.023 \pm 0.030$  \\
 $0.90$  &  $-0.093 \pm 0.028 \pm 0.075$  &  $-0.031 \pm 0.017 \pm 0.025$  &  $0.051 \pm 0.023 \pm 0.031$  \\
 $1.25$  &  $-0.017 \pm 0.020 \pm 0.059$  &  $-0.091 \pm 0.022 \pm 0.021$  &  $-0.012 \pm 0.017 \pm 0.034$  \\
 $1.89$  &  $-0.037 \pm 0.021 \pm 0.037$  &  $-0.090 \pm 0.026 \pm 0.023$  &  $0.014 \pm 0.017 \pm 0.042$  \\
 $3.11$  &  $-0.001 \pm 0.037 \pm 0.016$  &  $-0.243 \pm 0.055 \pm 0.063$  &  $0.032 \pm 0.027 \pm 0.062$  \\
\noalign{\smallskip}\hline\noalign{\smallskip}
\noalign{\smallskip}\hline\noalign{\smallskip}
$\langle x_F \rangle$ & $\lambda_\theta$ & $\lambda_\phi$ & $\lambda_{\theta \phi}$  \\
\noalign{\smallskip}\hline\noalign{\smallskip}
$-0.193$  &  $-0.092 \pm 0.037 \pm 0.067$  &  $-0.037 \pm 0.023 \pm 0.025$  &  $0.059 \pm 0.025 \pm 0.062$  \\
$-0.123$  &  $-0.117 \pm 0.026 \pm 0.050$  &  $-0.067 \pm 0.017 \pm 0.027$  &  $0.024 \pm 0.021 \pm 0.035$  \\
$-0.074$  &  $-0.070 \pm 0.024 \pm 0.043$  &  $-0.085 \pm 0.015 \pm 0.030$  &  $0.058 \pm 0.016 \pm 0.029$  \\
$-0.026$  &  $ 0.014 \pm 0.019 \pm 0.041$  &  $-0.112 \pm 0.023 \pm 0.034$  &  $0.073 \pm 0.016 \pm 0.033$  \\
 $0.021$  &  $-0.075 \pm 0.023 \pm 0.043$  &  $-0.100 \pm 0.027 \pm 0.040$  &  $0.042 \pm 0.020 \pm 0.048$  \\
 $0.067$  &  $-0.064 \pm 0.041 \pm 0.049$  &  $-0.028 \pm 0.029 \pm 0.047$  &  $-0.085 \pm 0.044 \pm 0.073$  \\
\noalign{\smallskip}\hline
\end{tabular}
\end{table*}

\begin{table*}
\centering \caption{Values of the parameter $\lambda_\theta$ measured in the
Collins-Soper frame as a function of the average reconstructed $p_T$ and $x_F$ for carbon
(C) and tungsten (W) data. The errors are statistical and systematic.}
\label{tab:results_CW}
\begin{tabular}{ccc}
\hline\noalign{\smallskip}
$\langle p_T \rangle$ (GeV/$c$) & $\lambda_\theta$ (C) & $\lambda_\theta$ (W)   \\
\noalign{\smallskip}\hline\noalign{\smallskip}
 $0.35$  &  $-0.38 \pm 0.10 \pm 0.11$     &  $-0.81 \pm 0.10 \pm 0.20$  \\
 $0.79$  &  $-0.271 \pm 0.062 \pm 0.078$  &  $-0.32 \pm 0.09 \pm 0.17$  \\
 $1.25$  &  $-0.192 \pm 0.054 \pm 0.066$  &  $-0.22 \pm 0.09 \pm 0.13$  \\
 $1.89$  &  $-0.147 \pm 0.062 \pm 0.052$  &  $-0.227 \pm 0.081 \pm 0.076$  \\
 $3.11$  &  $-0.164 \pm 0.049 \pm 0.039$  &  $-0.146 \pm 0.072 \pm 0.038$  \\
\noalign{\smallskip}\hline\noalign{\smallskip}
\noalign{\smallskip}\hline\noalign{\smallskip}
$\langle x_F \rangle$ & $\lambda_\theta$ (C) & $\lambda_\theta$ (W)   \\
\noalign{\smallskip}\hline\noalign{\smallskip}
$-0.193$  &  $-0.097 \pm 0.090 \pm 0.099$  &  $-0.32 \pm 0.12 \pm 0.08$  \\
$-0.123$  &  $-0.176 \pm 0.054 \pm 0.062$  &  $-0.18 \pm 0.15 \pm 0.10$  \\
$-0.074$  &  $-0.304 \pm 0.079 \pm 0.067$  &  $-0.29 \pm 0.07 \pm 0.10$  \\
$-0.026$  &  $-0.236 \pm 0.062 \pm 0.097$  &  $-0.37 \pm 0.09 \pm 0.10$  \\
 $0.021$  &  $ 0.03 \pm 0.15 \pm 0.15$     &  $-0.21 \pm 0.17 \pm 0.10$  \\
 $0.067$  &  $-0.71 \pm 0.31 \pm 0.23$     &  $-0.02 \pm 0.52 \pm 0.10$  \\
\noalign{\smallskip}\hline
\end{tabular}
\end{table*}

\section{Conclusions} \label{sec:conclusions}

HERA-B has measured the two-lepton decay angular distribution of $J/\psi$'s produced
inclusively in proton-nucleus collisions, using the decay channels $e^+ e^-$ and $\mu^+
\mu^-$. The distributions of the polar and azimuthal angles have been determined in three
different polarization frames. The results can be summarized as follows.
\begin{itemize}
  \item In the observed phase space, $\lambda_{\theta}$ is negative, indicating that the $J/\psi$'s
  are produced with a preferred spin component $0$ along the reference axis.
  \item There is a definite hierarchy for the values of the decay angular parameters
  measured in different frames: the polar and azimuthal parameters satisfy the relations
  \begin{eqnarray}
                |\lambda_{\theta}(\mathrm{HX})| & < & |\lambda_{\theta}(\mathrm{GJ})|
                < |\lambda_{\theta}(\mathrm{CS})| \nonumber \\
                |\lambda_{\phi}(\mathrm{HX})| & > & |\lambda_{\phi}(\mathrm{GJ})|
                > |\lambda_{\phi}(\mathrm{CS})|, \nonumber
  \end{eqnarray}
  while $\lambda_{\theta\phi}$ is significantly different from zero only in the GJ frame.
  \item The polarization effects depend on the kinematics of the $J/\psi$. In particular,
  the polar anisotropy increases with decreasing $p_T$ and is maximal
  in the limit $p_T \rightarrow 0$.
\end{itemize}

The different results obtained in the three frames -- in terms of both polar and
azimuthal distributions -- are an example which shows that an analysis limited to only
one frame and one polarization parameter is in general incomplete. For example, the
present measurement of only the polarization parameter $\lambda_{\theta}$ (i.e. ignoring
$\lambda_{\phi}$) in the HX frame for $p_T > 1$~GeV$/c$ may be misunderstood as a
significant indication of unpolarized $J/\psi$ production.

Among existing measurements of the parameter $\lambda_{\theta}$, E866~\cite{e866} (p-Cu
at $\sqrt{s} = 38.8$~GeV) has measured in the CS frame a $p_T$-independent polarization
consistent with zero, while the CDF Run II~\cite{cdf} data ($p\bar{p}$ at $\sqrt{s} =
1.96$~TeV) indicate a negative polarization in the HX frame increasing in magnitude with
increasing $p_T$. These results have been obtained in kinematic ranges (E866: $x_F >
0.25$, CDF: $p_T> 5$ GeV$/c$) which have no overlap with the HERA-B data and between each
other. The three results are therefore not in contradiction; their comparison has to be
interpreted as a further indication that the observed polarization effects change with
varying kinematic conditions of the produced $J/\psi$.

\subsection*{Acknowledgments} We express our gratitude to the DESY laboratory for the
strong support in setting up and running the HERA-B experiment. We are also indebted to
the DESY accelerator group for their continuous efforts to provide good and stable beam
conditions. The HERA-B experiment would not have been possible without the enormous
effort and commitment of our technical and administrative staff. It is a pleasure to
thank all of them.

\end{document}